\documentclass[manuscript]{aastex62}
\usepackage{graphicx}
\usepackage{epstopdf}

\shorttitle{The SMC Supernova Remnant Population}
\shortauthors{Leahy \& Filipovic}

\begin{document}


\title{Physical Properties of the Supernova Remnant Population in the Small Magellanic Cloud}


\author{D.A. Leahy}
\affil{Department of Physics $\&$ Astronomy, University of Calgary, Calgary,
Alberta T2N 1N4, Canada}
\author{M.D. Filipovi\'c}
\affil{School of Science, Western Sydney University, Locked Bag 1797, Penrith South DC, NSW 2751, Australia}



\begin{abstract}
{The X-ray emission from a supernova remnant is a powerful diagnostic of the state of its shocked plasma.
The temperature and the emission measure are related to the energy of the explosion, the
age of the remnant, and the density of the surrounding medium. 
Here we present the results of a study of the remnant population of the Small Magellanic Cloud.
Progress in X-ray observations of remnants has resulted in a sample of 20 remnants in the Small Magellanic 
Clound with measured temperatures and emission measures.
We apply spherically symmetric supernova remnant evolution models to this set of remnants, 
to estimate  ages, explosion energies, and circumstellar medium densities.
The distribution of ages yields a remnant birthrate of $\sim$1/1200 yr. 
The energies and densities are well fit with log-normal distributions, with means of 1.6$\times10^{51}$ erg and 0.14 cm$^{-3}$, and 1$\sigma$ dispersions of a factor of 1.87 in energy and 3.06 in density, respectively.
}
\end{abstract}



\section{Introduction}

The study of supernova remnants (SNRs) is important for astrophysics, including the understanding of injection 
of energy (e.g. \citealt{2005Cox}) and elements into the interstellar medium (e.g. \citealt{2012Vink}). 
Valuable constraints for stellar evolution and the evolution of the interstellar medium and the Galaxy can
be obtained from SNR studies. 
The goals of SNR research include understanding explosions of supernovae (SN) and the resulting energy 
and mass injection into the interstellar medium.

Approximately 300 SNRs have been observed in our Galaxy (e.g. \citealt{2012Vink}).
Several historical SNRs have 
been 
modelled with hydrodynamic simulations (e.g. \citealt{2006ApJ...645.1373B}).
For many other SNRs, it has been customary to apply a simple Sedov model with an assumed energy and ISM density.
Better modelling is required to derive reliable energies, densities and ages from observations.
A set of SNR models for spherically symmetric SNRs, which are based on hydrodynamic calculations have been
developed by \citet{2019AJ....158..149L}. 
These models calculate emission measures (EM) and EM-weighted temperatures for forward-shocked and for 
reverse-shocked gas.
The resulting SNR models allow the use of X-ray observations to estimate SNR physical properties.
43 Galactic SNRs with X-ray spectra and distances were modeled in this way by \citet{2020ApJS..248...16L} with resulting distributions of ISM density, explosion energy and SNR ages derived for this set. 
This work showed, not surprisingly, that the range of energies and densities was large.
The distributions densities and explosion energies were found to be well-described by log-normal functions. 
The age distribution was fit to obtain the SNR birthrate. 

Our nearby galaxies, the Large and Small Magellanic Clouds (LMC and SMC) are at well known distances of
 50 kpc and 60 kpc, respectively ( \citealt{2019Natur.567..200P} ;  \citealt{2005MNRAS.357..304H}; \citealt{2014ApJ...780...59G}).
They are close enough that we can detect and spatially resolve
SNRs from radio to X-ray wavelengths, and are located behind
only a moderate Galactic foreground (N$_{H}$ of a few $10^{20}$ cm$^{-2}$; 
\citealt{1990ARA&A..28..215D}). Models for the SNR population of the LMC and their ages, densities and energies,
were presented by \cite{2017Leahy}. 

Here we study the SNR population of the SMC. This work includes the following. 
The sample of 19 SMC SNRs with measured X-ray EM and kT from \cite{2019AA...631A.127M} is used, 
supplemented by SNR E0102-7219.
 SNR E0102-7219 is the brightest SNR in the SMC, and has been extensively studied in X-rays (e.g.  
 \citealt{2020ApJ...904...70L}, \citealt{2019ApJ...873...53A}, \citealt{2001A&A...365L.237S} \citealt{2001A&A...365L.231R}).
We apply the models of \citet{2019AJ....158..149L}, using the solution to the inverse problem as
described in  \citet{2020ApJS..248...16L} to obtain SNR best fit properties and their uncertainties.
The structure of the paper is as follows.
In Section~\ref{sec:datamodel} we describe the sample of SMC SNRs then provide an overview of the SNR models and 
the application to the SMC SNR sample. 
The results of model fits are given in Section~\ref{sec:SNRmodels}.
In Section~\ref{sec:properties} we discuss the results, including comparison of statistical properties of the SMC SNR population with the LMC and Galactic SNR populations.
The conclusions are summarized in Section~\ref{sec:conclusion}.

\section{Data and Model Description}\label{sec:datamodel}

\subsection{The SNR Sample}\label{sec:sample}

For modelling we assume a common distance of 60 kpc to all SNRs
and take SMC ISM abundances from column 1 of  Table 3 in  \cite{2019AA...631A.127M}.
The basic parameters of the 19 SNRs were obtained from Table A.3 of \cite{2019AA...631A.127M}.
The SMC SNR E0102-7219 (MCSNR J0104-7201) is not included in the X-ray spectral analysis of \citet{2019AA...631A.127M}.
It is the only one of the SMC SNRs of Table A.1 of \citet{2019AA...631A.127M} that is not included in their spectral analysis,
thus we obtain its X-ray spectral parameters from \citet{2019ApJ...873...53A}.
For each SNR, the EM and EM-weighted temperature were derived from the X-ray studies.
The average of semi-major and semi-minor axes from  Table A.1 in \cite{2019AA...631A.127M} 
 was used as the outer shock radius for input to the spherical SNR model. 
 
For SNRs which are faint in X-rays, which applies to most of the SMC SNRs, the classification as CC type or Ia type depends on whether the
X-ray spectrum shows enhanced O, Ne and Mg (CC-type) or enhanced Fe (Type Ia).
The Fe enhancements in the X-ray spectra of  IKT 5 (SNR) J0047-73.14), IKT 25 (SNR J0104-7205), and DEM S 128 (SNR J0105-7210) 
are used by \citet{2019AA...631A.127M} to classify these SNRs  as Type Ia. 
However \citet{2015ApJ...803..106R} raise questions about the Type Ia classification of these three SNRs, suggesting that they may be CC type SNRs.
They point out that although they are overabundant in Fe, they show other features, including potential point sources, X-ray asymmetry, and radio optical and infrared features that are not typical of Type Ia SNRs. Thus 
we mark these three SNRs as uncertain Type Ia by ``Ia?''.

\subsection{The Model for SNR Evolution}\label{sec:model}

A SNR is the interaction of the SN ejecta with the circumstellar/interstellar medium (CSM/ISM). 
The various stages of evolution of a SNR are labelled the 
ejecta-dominated stage (ED), the adiabatic or Sedov-Taylor stage
(ST), the radiative pressure-driven snowplow (PDS) and the radiative momentum-conserving shell (MCS).
These stages are reviewed in, e.g., \cite{1988cioffi}, \cite{1999truelove} (hereafter TM99), 
\citet{2012Vink} and \citet{2017LeahyWilliams}.
In addition, there are the transitions between stages, called ED to ST, ST to PDS, and PDS to MCS, respectively.
The ED to ST stage is important because the SNR is still bright, and it is long-lived enough that a significant 
fraction of SNRs are likely in this phase. For example, a type II SNR with ejecta mass of 10 M$_{\odot}$ and average 
explosion energy of $5\times10^{50}$ erg \citep{2020ApJS..248...16L} in an ISM density of 0.1 cm$^{-3}$, is in the 
ED phase up to age $\sim$4000 yr, and in the ED to ST transition phase until age $\sim$23000 yr \citep{2019AJ....158..149L} . 

For simplicity our models assume that the SN ejecta and CSM/ISM are spherically symmetric.
The CSM/ISM density profile is a power-law centred on the SN explosion, given by $\rho_{ISM}=\rho_s r^{-s}$, 
with s=0 (constant density medium) or s=2 (stellar wind density profile).
The unshocked ejectum has a power-law density  $\rho_{ej}\propto r^{-n}$.
With these profiles, the ED phase of the SNR evolution has a
self-similar evolution (\citealt{1982Chev} and \citealt{1985Nad}).
The evolution of  SNR shock radius was extended for the ED to ST and ST phases by TM99.

A detailed description of the SNR evolution model is given in \citet{2019AJ....158..149L} and \citet{2017LeahyWilliams}. 
The forward model, using initial SNR conditions and calculating conditions at time t, is described in \citet{2019AJ....158..149L}.
The input parameters of the SNR forward model are: age;
SN energy $E_0$; ejected mass $M_{ej}$; ejecta power-law index n; ejecta composition;
ISM temperature $T_{ISM}$; ISM composition; ISM power-law index s (0 or 2); and
ISM density $n_0$ (if s=0) or mass-loss parameter $\rho_s=\frac{\dot{M}}{4\pi v_w}$ (if s=2).
The method of application to observations of SNRs is given by \citet{2020ApJS..248...16L}.
Essentially the SNR evolution based on  the initial conditions of the explosion is inverted so that the initial
quantities are calculated from the SNR current observed properties. 
The inverse problem is solved by an iterative procedure to obtain a unique solution
for density, energy and age for specified values of $s$ and $n$. 
 
 \subsection{Application of Models to the Sample}\label{sec:application}

We apply two sets of models  to the SNR sample. 
The first set of models is applied to the full sample of   
20 SNRs. 
We fit the data with a single hot plasma component, which is taken to be emission from gas heated
by the forward shock (FS) and set $s=0$, $n=7$ .
The outer part of the envelope of Type Ia progenitors has a density profile with $n=7$
\citep{1969ApJ...157..623C}. For Type II/CC progenitors, the envelopes may have steeper 
radial profiles \citep{1982ApJ...259..302C}, with values of $n$ from 7 to 12.5.
For the Type Ia? or unknown type SNRs we set the ejecta mass $M_{ej}$=1.4$M_{\odot}$ and   
 for the CC SNRs we set $M_{ej}$=5 $M_{\odot}$. This is called the Standard Model.
For E0102-7219, the inferred progenitor mass is 40 $M_{\odot}$ \citep{2019ApJ...873...53A},
thus we run additional models with ejected mass of 20 and 40  $M_{\odot}$: 
the first corresponds to a 50\% mass loss prior to explosion, and the second to no mass loss. 
The values for those cases are added as a footnote to Table~\ref{tab:TBLstdmodel}, and yield
a $\sim$15\% and $\sim$30\% increase in age; a $\sim$7\% and $\sim$10\% decrease in energy;
and a $\sim$6\% and $\sim$12\% decrease in density, respectively.

The second set of models is carried out for those 
6 SNRs with two measured hot plasma components for the whole SNR: 
5 from \citet{2019AA...631A.127M}
and  E0102-7219 from \citet{2019ApJ...873...53A}.
The two components are from the forward shocked gas and the reverse shocked gas, with the reverse shocked gas identified by its enhanced element abundances.
For the second set of models, $R_{FS}$, $EM_{FS}$ and $kT_{FS}$ as taken as model inputs and we compute predicted
$EM_{RS}$ and $kT_{RS}$ for different $s$, $n$ cases to test which $s$, $n$ best describe the data. 

\section{Results of SNR Models}\label{sec:SNRmodels}

The results from the Standard Model model are shown in Table~\ref{tab:TBLstdmodel}.
Derived ages, energies and densities and their uncertainties are given. 
Uncertainties were found by running models with the input parameters set to their upper and lower limits in different combinations and choosing the largest deviation from the results for the best-fit input parameters.

\begin{figure*}[ht!]
\plottwo{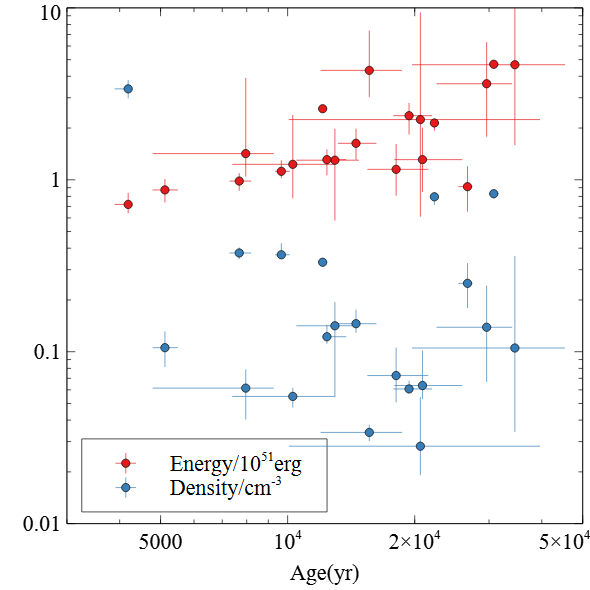}{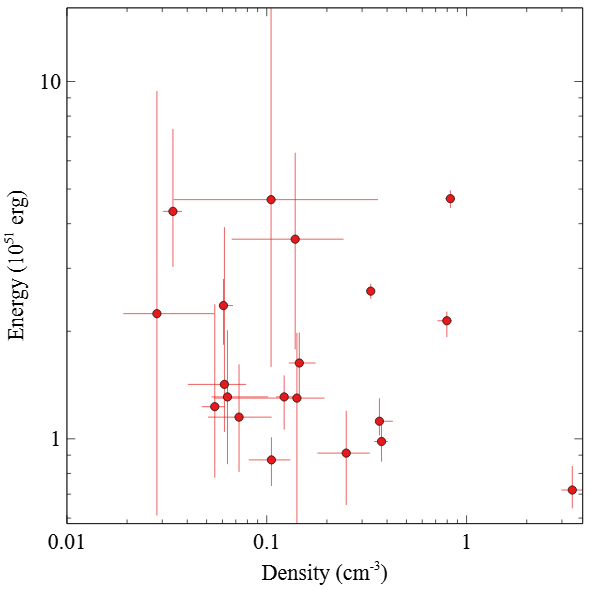}
\caption{Left panel: Energy and density vs. age for the  
20 SNRs from the standard model (forward shock from an SNR with ISM and ejecta profiles with $s=0$, $n=7$). Right panel: Energy vs. density for the 20 SNRs 
from the standard model.}
\label{fig:stdmodel}
\end{figure*}

The left panel of Figure~\ref{fig:stdmodel} shows the derived explosion energies and ISM densities vs. SNR ages,
and the right panel shows the explosion energies vs. ISM densities.
The SNR ages range from  $\sim$4000 yr to 35000 yr, the explosion energies range from $\sim$0.7 to 5$\times10^{51}$ erg,
and the ISM densities range from $\sim$0.03 to 3 cm$^{-3}$.
The figures illustrates the wide range of all three parameters and lack of correlation between them for the SNR sample.
Analysis of the results of the Standard Model are discussed in Section~\ref{sec:snrstats} below.

We apply our model to the 6 SNRs with a  second X-ray emission component for 6 different cases of (s,n): (0,7), (0,10), (2,6),  (2,7),  (2,8) and (2,10).
The $s$=0 cases are for a uniform circumstellar medium and $s$=2 cases are for a stellar wind density profile. 
Type Ia could be either, depending on whether or not the white dwarf has a companion with a strong stellar wind. 
CC SNRs can be either, depending on the presence or not of a strong stellar wind for a long enough time prior to explosion. 
More $n$ values are chosen for the $s$=2 case because the effect of changing $n$ is stronger for $s$=2 than for $s$=0 \citep{2019AJ....158..149L}.
The model returns the age, energy and density required in order to reproduce the forward shock.
We calculate the properties of the reverse shock  ($EM_{RS}$ and $kT_{RS}$) from the model for these different cases
and give the results in Table~\ref{tab:customS2}.
 Table~\ref{tab:customS2} also gives the observed values of $EM_{RS}$ and $kT_{RS}$ obtained from fitting the observed X-ray spectrum.

The reverse shock EM increases as $n$ increases, and is larger for $s$=2 than $s$=0
(see  Table~\ref{tab:customS2}).
The reverse shock kT is larger for s=0 than for s=2. 
With the exception of (s,n)=(0,6), the model EM-weighted reverse shock temperature decreases as n increases.
The results from modelling the two-component SNRs are discussed in Section~\ref{sec:2comp} below.

\subsection{Uncertainties and systematic effects}\label{sec:systematics}

The uncertainties are described briefly here and are described in detail in \cite{2020ApJS..248...16L}.
Calibration uncertainties of the X-ray different instruments used to measure $EM$ and $kT$ 
are typically lower than $\sim$10\%. 
EM errors are typically 10 to 50\%, and kT errors typically 5 to 50\%, so usually dominate over instrument uncertainties.

\cite{2020ApJS..248...16L} considered the effect of changing ISM abundances on the model output parameters. 
I.e., for given observed conditions, the SNR has a X\% decrease/increase in parameter B (age, density or explosion energy).
The high abundance case (factor of $10^{0.5}$ above solar) yielded a $\sim 1.2$\% increase in age, and low abundance 
(factor of $10^{0.5}$ below solar) yielded  a $\sim 0.4$\% decrease in age.
For ISM density, the change is systematic with high abundance yielding a $-0.7$\% change and
low abundance yielding a $+0.25$\% change.
For  explosion energy, the high abundance case yields a $\sim 1.5$\% decrease, and low abundance yields  
a $\sim 0$ to $\sim0.8$\% increase.
In suumary all output values change less than 2\% as ISM abundance is changed.

The effect of changing the ejecta masses is as follows. 
For Type CC SNRs, increasing ejecta mass from 5 to 10$M_{\odot}$ increased the age by 0 
to 20\%, with mean change of 4\%.
Increased ejecta mass decreases the ISM density by 0 to 7\%, with mean decrease of 2\%.
The explosion energy had a 9\% decrease to a 42\% increase, with mean increase of 18\% and large standard deviation of 20\%.
For Type Ia SNRs, decreasing ejecta mass from 1.4 to 1.0$M_{\odot}$ decreased the age by 0 
to 6\%, with mean decrease of 2\%.
The decreased ejecta mass changed the ISM density by -1 to +3.5\%, with mean increase of 1.5\%.
The explosion energy for Type Ia had a change of  -1 to +4.5\% increase, with mean increase of 2\%.
Overall, the change in mean age and mean ISM density for all cases is small (between 1\% and 4\%),
 whereas the change in mean explosion energy is moderate (between 1\% and 20\%).
 
\section{Discussion}\label{sec:properties}

\subsection{Standard Model}\label{sec:snrstats}

\subsubsection{Statistical properties of SMC SNRs}\label{sec:stats}

The statistical properties of the ages, explosion energies and ISM densities of the SNR sample are examined
using an analysis similar to that carried out for the LMC SNR population \citep{2017Leahy}, 
for 15 Galactic  SNRs \citep{2018ApJ...866....9L} and for an additional 43 Galactic  SNRs \citep{2020ApJS..248...16L}.

The model age distribution gives information on the SN birth rate.
As noted above, the SMC SNR E0102-7219 (MCSNR J0104-7201) is not included in the X-ray spectral analysis of \citet{2019AA...631A.127M}. 
\citet{2021ApJ...912...33B} use proper motion measurements of oxygen-rich knots to find an age of E0102-7219 of 1738$\pm$175 yr.
However \citet{2019ApJ...873...53A} find a Sedov age of 3500 yr\footnote{We have confirmed that  age by running a Sedov model for E0102-7219 with their shocked ISM temperature and emission measure. This is a lower limit to age, because adding a non-zero ejecta mass increases the age.}.
Table~\ref{tab:TBLstdmodel} uses our model calculated parameters\footnote{We have also used an alternate set of ages with age of E0102-7219 
set to 1738$\pm$175 yr and verified that did not affect the birthrate results here within the errors.}. 

\begin{figure*}[ht!]
\plottwo{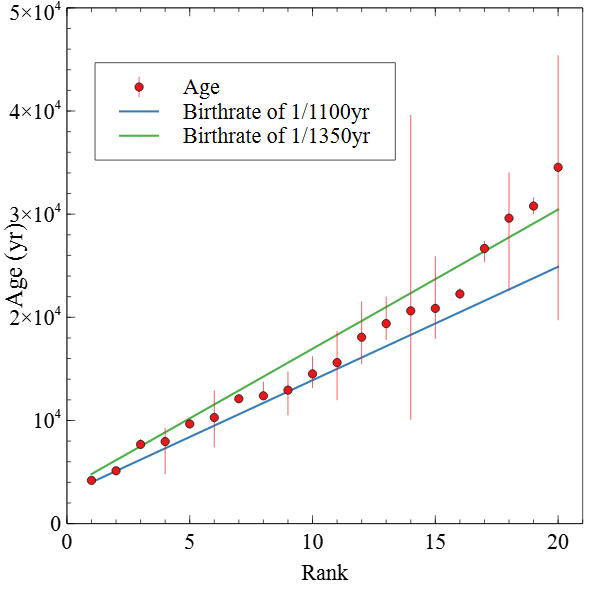}{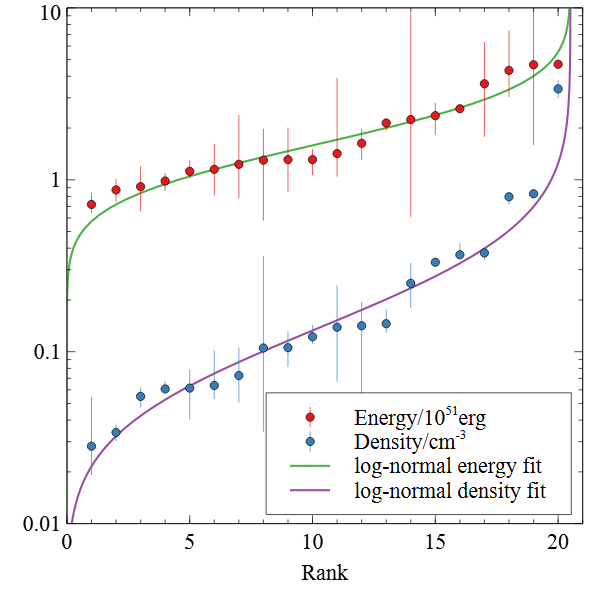}
\caption{Left panel: Cumulative distribution of model ages from the standard model, and fit lines for 
birth rates of of 1 per 1200 yr and 1 per 1350 yr. 
For each parameter of interest (age, energy or density), the sample was sorted in increasing order of parameter to obtain the rank
for that parameter. 
The rise in age with rank for the last 4 SNRs is probably caused by incompleteness.
Right panel: Cumulative distributions of s=0, n=7 model  explosion energies and of densities of the 20 SNRs. The fit lines are best fit log-normal distributions. 
For explosion energies the mean is E$_0$=1.6$\times10^{51}$ erg and the 1$\sigma$ dispersion in log(E$_0$) is 0.271  (a factor of 1.87).
For densities the mean is $n_0$=0.14 cm$^{-3}$ and  the 1$\sigma$ dispersion in log($n_0$) is 0.486 (a factor of 3.06).
}
\label{fig:ageE51D}
\end{figure*}

We add this SNR to the age distribution and show the age distribution in the left panel of  Figure~\ref{fig:ageE51D}. %
For each parameter of interest (age, energy or density), the sample was sorted in increasing order of parameter to obtain the rank
for that parameter. 
The straight lines are the expected distributions for constant birth rates of 1/(1100 yr) yr and  1/(1350 yr), respectively. 
	The rise in age with rank for the last 4 SNRs is probably caused by incompleteness, which means the true rank (rank if the missing SNRs were included) of those 4 is larger. 
Thus when we fit for birthrate we omit those 4 from the sample.
The lines in Fig.~\ref{fig:ageE51D} have non-zero x-intercepts of -2.64 and -2.58, 
respectively.\footnote{This is equivalent to allowing for $\sim$2 to 3 missing young (age $\lesssim$ 4000 yr) SNRs in the sample.}

 A birthrate of 1/(1230 yr) line is the best fit to the youngest 16 SNRs of the 20 SNR sample, where the completeness is best, 
with $\chi^2=$10.9 for 14 degrees of freedom. With 2 parameters of interest (slope and intercept of the line) the 95\% (2 $\sigma$) confidence 
interval (e.g \citealt{2002nrca.book.....P}) on birthrate is 1/(1100 yr) to 1/(1350 yr).
The non-zero x-intercept of the fits suggest there are $\sim$2-3 missing young SNRs in our sample.
Alternately, the stochastic behaviour of SN explosions could explain the lack of $\sim$2-3 young SNRs.
 
The cumulative distribution of explosion energies is shown in  the right panel of  Figure~\ref{fig:ageE51D}. 
This distribution is consistent with a log-normal distribution. 
The property that SNR explosion energies follow a log-normal distribution was discovered by \citet{2017Leahy} for LMC SNRs
and verified for Galactic SNRs by  \citet{2018ApJ...866....9L} and  \cite{2020ApJS..248...16L}. 
Here, the best fit 
of a cumulative log-normal distribution to the observed
cumulative distribution has an average energy (peak of the log-normal)
$E_{av}= 1.6\times10^{51}$ erg and dispersion $\sigma_{logE}$=0.265 (a 1-$\sigma$ dispersion factor for $E$ of 1.87).

 The right panel of  Figure~\ref{fig:ageE51D} shows the cumulative distribution of ISM densities.
This distribution is consistent with a log-normal distribution, similar to that found previously for the LMC and the Galaxy.
The best fit cumulative distribution is shown by the solid line. 
The average density is $n_{0,av}= 0.14$ cm$^{-3}$ and dispersion is $\sigma_{log(n_0)}$=0.486 ( a 1-$\sigma$ dispersion factor for $n_0$ of 3.06). 
A log-normal distribution is the natural resulting probability distribution if a physical process is caused by a set of many random processes, 
 where the net physical process depends on the product of the individual processes. Thus both the explosion energy distribution and the 
 ISM density distribution are inferred to be caused by a set of individual processes which act as a product (i.e., in series).
 
\begin{figure*}[ht!]
\includegraphics[width=0.6\textwidth]{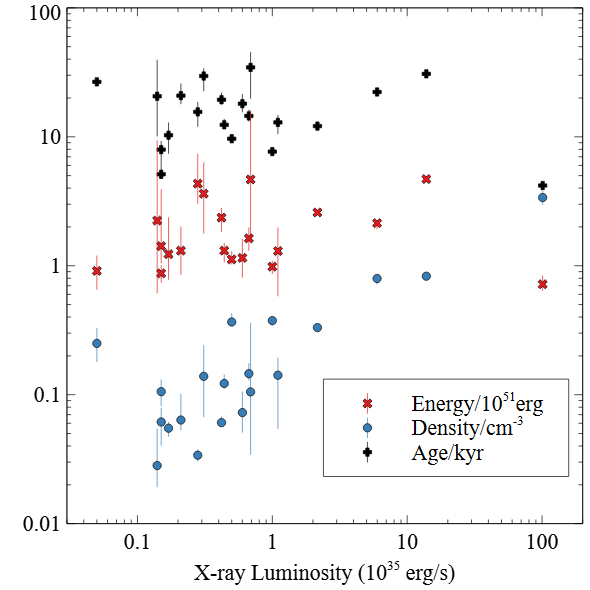}
\caption{
Standard model energy, density and age vs. observed X-ray luminosity for the 20 SNRs.}
\label{fig:Dhist}
\end{figure*}

 The correlation of observed X-ray luminosity with model parameters (energy, density and age)
 is shown in Figure~\ref{fig:Dhist}. 
 The explosion energy and age show no correlation with X-ray luminosity. 
 However the ISM density and luminosity are strongly correlated. 
 This is because the normalization of FS emission depends on its emission measure which scales as square of FS density times emission volume.
 From Figure~\ref{fig:Dhist}, the correlation of density ($n_0$) on X-ray luminosity ($L_X$) follows a $\sim$2/3 powerlaw, so luminosity depends on density as $L_X\propto n_0^{3/2}$.
 This is weaker than the expected $n_0^2$ probably because of other effects.
  These other effects are seen in the left panel of Figure~\ref{fig:stdmodel}.
  ISM density is negatively correlated with age and SNR energy is weakly but positively correlated with age.
 The former is likely caused by the shorter lifetimes of SNRs in high density ISM
 and could weaken the correlation of X-ray luminosity and density, as seen in  Figure~\ref{fig:Dhist}.

\subsubsection{Comparison with LMC and Galactic SNRs}\label{sec:compLMCGal}

Table~\ref{tab:stats} compares the results for SNR birthrate, energy and ISM density from the current work with the
values for the LMC and the Galaxy from previous work.
The birthrates are expected to be different because of the different total stellar masses of SMC, LMC and the Galaxy, and different
selection (incompleteness) effects for the three systems. Because the extinction is low for the SMC and LMC the X-ray samples of SNRs
are expected to be complete down to a given X-ray luminosity. The incompleteness for the Galaxy is expected to be higher and
have strong spatial dependence because of the high extinction in the Galactic plane, particularly for SNRs further from the
Sun. An estimage of the observed SNR birthrate for the Galaxy can be obtained by combining the 15 SNR and 43 SNR samples, which have no SNRs in common.
The result is an observed birthrate of 1/(236 yr). Correction for selection effects is beyond the scope of the current work. 
However, this is lower than the SN rate of 1/(40 yr) for our Galaxy, including all types of SN \citep{1994Tammann},
which yields an incompleteness in the 58 SNR sample of a factor $\sim$6.

$E_{av}$ for the SMC SNRs is a factor of 3-4 times larger than the values from the LMC SNR sample or from the two Galactic SNR samples. 
That the SMC SNR energies are significantly higher is not related to sensitivity of the observations. 
The LMC is nearer at 50kpc than the SMC at 60kpc so a given flux lower limit results in a luminosity lower limit 1.44 times larger for the SMC. 
The LMC and SMC surveys did not have a uniform set of exposure times, although the coverage of the SMC survey was 
significantly more complete than that of the LMC (\citealt{2016AA...585A.162M},\citealt{2019AA...631A.127M}).  
However, the faintest SNRs detected in both SMC and LMC have 0.3-8 keV X-ray luminosities of 5-7$\times10^{33} $erg/s  \citep{2022arXiv220110026F}.
The higher energy may be related to the lower metallicity of the SMC than either the LMC or the Galaxy:
lower metallicity stars have weaker stellar winds, thus result in higher stellar mass at the time of explosion for given initial stellar mass.
It may also be related to a recent burst of star formation which yielded energetic-SN progenitors. 

The mean ISM density of the SMC SNR sample is higher than for the LMC SNRs,
and the dispersion is smaller. The lower metallicity of SMC progenitor stars is expected to result in weaker stellar winds,
which would lead to an ISM which is less affected by large low density stellar wind bubbles, and thus have a higher mean density
and also a lower dispersion in density.
Supporting this argument, the dispersion in ISM density for the Galaxy is higher than either SMC or LMC, which is expected from
the more inhomogeneous environments in the Galaxy\footnote{Galactic SNRs occur over a wide range of Galactocentric distances, 
thus large change in mean Galactic disk density, and higher metallicity for Galactic stars is expected to result in 
stronger stellar winds than for SMC or LMC stars.} compared to SMC or LMC.
Combining the 15 Galactic SNRs in the inner Galaxy (l=18-54$^\circ$) with the 43 Galactic SNRs, we find a mean density of 
0.12 cm$^{-3}$. 
The mean density should approximately reflect the density at the mean Galactocentric radius of Galactic SNRs. 

\subsection{Two-component SNRs}\label{sec:2comp}
Six of the SNRs (J0049-7314, J0051-7321, J0105-7210,  J0104-7201 (=E0102-7219), J0105-7223 and J0106-7205) have two measured X-ray emission components, from forward-shocked (FS) and reverse-shocked (RS) material.
For J0104-7201, the whole-SNR hot plasma components are calculated from the sub-regions given by
\citet{2019ApJ...873...53A}\footnote{We calculate the total FS emission measure of
8.6(+2.3,-1.9)$\times10^{59}$cm$^{-3}$ from their value for their Shell region, extended to cover the whole SNR. We verified their SNR Sedov age of 3500 yr using that value. 
The RS emission measure and temperature, shown in Table~\ref{tab:customS2} here, were obtained from their 178 different
A to G regions (their Tables 5 to 12 and Figure 5). The total EM was the sum of EM values corrected for areas not covered. The EM-weighted RS temperature was obtained using EM and kT values for all shocked ejecta regions.}.
We  applied (s,n)= (0,7),  (0,10), (2,6), (2,7), (2,8) and (2,10) models by requiring each model to match the measured FS quantities,
$R_{FS}$, $EM_{FS}$ and $kT_{FS}$. 
The results are shown in Table~\ref{tab:customS2}.

Given the assumptions of the SNR models, we assess their ability to reproduce the observed $kT_{RS}$ and $EM_{RS}$ of the SMC SNRs
with observed $RS$ properties. 
The models listed in Table~\ref{tab:customS2} were calculated assuming CC abundances from \citet{2017LeahyWilliams}.
For CC-type, we use $M_{ej}=5M_{\odot}$; for Ia?-type, we use $M_{ej}=1.4M_{\odot}$. 
We approximately take into account the effects of ejecta abundances (Ia vs. CC) on $EM_{RS}$ and $kT_{RS}$ using the 
results of abundance and partial ionization effects from Leahy (2022, in preparation).

Including these effects, 
the change in $EM_{RS}$ is by factors between 0.01 to 1.5 and the change in $kT_{RS}$ is by factors between 0.1 and 2,  
with the largest decrease in both $EM_{RS}$ and $kT_{RS}$ caused by changing abundances from CC-type to Ia-type.

\subsubsection{Two-component models for individual SNRs}\label{sec:individ}

\textbf{J0049-7314/Ia?}: Of the given CC models, (s,n)=(2,8) is closest with $EM_{RS}$ consistent within errors 
but $kT_{RS}$ too low by factor 0.22. This SNR is Type Ia?, which decreases the model $EM_{RS}$ and $kT_{RS}$ values, making the disagreement worse. The  (s,n)=(0,7) model has a Type Ia adjusted
$kT_{RS}$ of 0.76$\pm0.23$ keV, consistent with the data, but far too small  $EM_{RS}$.  
A possibility that this SN has a significant amount of light elements, including hydrogen, mixed in the ejecta, increasing the model $EM_{RS}$, 
and NEI effects raise the model $kT_{RS}$ to be consistent with the data.

\textbf{J0051-7321/CC}:  None of the listed CC models match the data. 
Model (s,n)=(2,7) has the closest values, a factor 3 to 4 higher (for $EM_{RS}$) and lower (for $kT_{RS}$) than the data.
Fewer light elements in the ejecta can lower the model $EM_{RS}$ to match the data, and NEI can raise $kT_{RS}$ to be less discrepant.

\textbf{J0104-7102/CC}: 
The measured $kT_{RS}$ and $EM_{RS}$ are between those models with (s,n)=(2,7) 
and (s,n)=(2,8), thus this SNR is consistent with (s,n)=(2,$\sim$7.5). 

\textbf{J0105-7210/Ia?}: CC model (s,n)=(2,7) is  consistent with both measured $kT_{RS}$ and $EM_{RS}$. 
However this SNR is classified as Type Ia?, so the model values should be adjusted down and become not consistent with the data. 
We conclude that this SN may have a significant amount light elements 
mixed in the ejecta which keeps $kT_{RS}$ and $EM_{RS}$ close to the CC model values.

\textbf{J0105-7223/CC}: CC model (s,n)=(2,7) is the closest one to the measured $kT_{RS}$ and $EM_{RS}$ (within factors of $\sim$1.2), thus this model is consistent with the SNR.

\textbf{J0106-7205/Ia?}:  The (s,n)=(2,7) model is the closest one to the measured $kT_{RS}$ and $EM_{RS}$ (within factors of $\sim$1.5).
Similar to the case of J0049-7314, this SNR is classified as Type Ia?, so we conclude  this SN may have a significant 
amount light elements mixed in the ejecta.

\subsubsection{Discussion of two-component models}\label{sec:individ}

Three of the six 2-component SNRs are classified as Ia?, and three as CC. 
Above we applied a standard set of CC abundances for the composition of the ejecta, which was assumed to be uniformly mixed. 
Use of uniformly mixed Type Ia abundances, with little hydrogen in the ejecta, yields smaller EM for RS material, because EM is defined in terms of hydrogen abundance.
This makes the calculated EM very sensitive to the assumed hydrogen abundance (see Section~\ref{sec:2comp}). 
Thus we chose to use CC abundances and estimate the correction to Type Ia abundances after calculating the SNR models.
For the three Type Ia? SNRs  the SNR type classification might be in in error \citep{2015ApJ...803..106R}.
However we think the sensitivity of the EM of RS material to the assumptions of RS composition is more likely what is making the comparison of models with data difficult to interpret.

Two SNRs (J0104-7102/CC and J0105-7223/CC) are well matched by the CC model for $kT_{RS}$ and $EM_{RS}$.
For the other four 2-component SNRs,
effects of changes in ejecta composition and of partial ionization 
considered by Leahy (2022, in preparation) are needed to change the model
$kT_{RS}$ and $EM_{RS}$ to be consistent with the data.   
The models are consistent with the three Type Ia SNRs if they have a significant amount of hydrogen mixed into the ejecta. 
The CC-type SNR J0051-7321 is consistent with the models if the amount of light elements in the ejecta is moderately less than for the standard assumed CC abundances.
A possible alternate solution to the $EM_{RS}$ and $kT_{RS}$ discrepancies is to have inhomogeneous ejecta.
These effects could be investigated by applying individual complex hydrodynamic solutions including inhomogeneities. 

One tentative result is that the best-estimated models for the six two component SNRs are 
in stellar wind environments rather in uniform density environments.
The previous study on Galactic SNRs by \cite{2020ApJS..248...16L}  had a similar conclusion:
 i.e. most two-component SNRs occur in a circumstellar wind environment. 
In that study, nine of the twelve  two-component SNRs were found to be more consistent with stellar wind environments and only three more consistent with a uniform circumstellar medium. 
More work on the effects of abundance variations, partial ionization and deviations from spherical symmetry needs to be
carried out to verify whether there is a large fraction of two-component SNRs in stellar wind environments. 
Because SNRs in a stellar wind have significantly larger $EM_{RS}$ than those in a uniform ISM (e.g., see Table~\ref{tab:customS2} here, 
where the s=2 SNRs have $EM_{RS}$ $\sim$1000 times larger than s=0 SNRs), 
 two-component SNRs should be strongly biased to detection of stellar wind SNRs.
This means that the high fraction of stellar wind SNRs in the two-component SNR sample, does 
not likely extend to the full sample of SNRs. 

\section{Conclusion}\label{sec:conclusion}

We utilize recent X-ray observations of 20 SNRs in the SMC. 
The  SNR models of \citep{2019AJ....158..149L} are applied to these SNRs with inputs their observed radii, emission measures and temperatures.
The models yield explosion energies, circumstellar densities and ages.
The distributions of the parameters were used to estimate properties of the SMC SNR population.
The energies and ISM densities of SNRs can be well fit with log-normal distributions in agreement
with our earlier studies (\citealt{2017Leahy}, \citealt{2018ApJ...866....9L}, \citealt{2020ApJS..248...16L}).
The mean density of SNR environments in the SMC is 0.1 cm$^{-3}$. 
The mean explosion energy is $1.6\times10^{51}$ erg, which is significantly higher than the mean energy of LMC
SNRs  ($\simeq5\times10^{50}$ erg or of Galactic SNRs  $\simeq4\times10^{50}$). 
The derived birthrate of SMC SNRs is $\simeq$1 per 1200 yr

Six of the 20 SNRs have two component X-ray spectra, i.e. both forward shocked material and
reverse shocked ejecta are detected. 
We apply different (s,n) models to these SNRs and conclude that other effects not built into the SNR models must be 
important for the models to match both forward shock and reverse shock properties.
These effects include changes in ejecta composition and partial ionization, discussed quantitatively
by Leahy (2022, in preparation).
By accounting for these effects, the best models for these SNRs are likely models for a SNR in a stellar wind environment. 
Further exploration of two-component SNRs requires more complex model calculations which include the effects above 
or non-spherical models.
 
\acknowledgments
This work was supported by a grant from the Natural Sciences and Engineering Research Council of Canada.

\clearpage



\clearpage

\clearpage

\begin{deluxetable}{crrrrrr}
\tabletypesize{\scriptsize}
\tablecaption{All SNRs: Standard (s=0, n=7 forward shock) Model Results\label{tab:TBLstdmodel}} 
\tablewidth{0pt}
\tablehead{\\
\colhead{SNR} & \colhead {SN type} & \colhead {$M_{ej}$} & \colhead{Age(+.-)} &  \colhead{Energy(+,-)}    & \colhead{Density(+,-)}   \\
\colhead{}   & \colhead{}   & \colhead{($M_{\odot}$)} & \colhead{(kyr)}  & \colhead{($10^{50}$ erg)}  & \colhead{($10^{-2}$ cm$^{-3}$)} 
}
\startdata
J0041-7336	&	CC 	& 	5	&	19.4(+2.6,-1.6)	&	23.6(+4.4,-5.3)	&	6.1(+0.7,-0.3) \\
J0046-7308	&	CC 	& 	5	&	14.5(+1.7,-1.4)	&	16.3(+3.5,-3.3)		&14.6(+3.0,-1.7) \\
J0047-7308	&	CC 	& 	5	&	9.7(+0.5,-0.3)	&	11.2(+1.8,-1.0)	&	36.6(+6.1,-0.6) \\
J0047-7309	&	CC 	& 	5	&	12.9(+1.8,-2.5)	&	13.0(+6.8,-7.2)		&14.2(+5.3,-8.7) \\
J0048-7319	&	unknown 	& 	1.4	&8.0(+1.3,-3.2)	&	14.2(+24.9,-3.8)	&	6.1(+1.7,-2.1) \\
J0049-7314	&	Ia? 	&  	1.4	&	18.1(+3.5,-2.6)	&	11.5(+4.7,-3.4)		&7.3(+3.3,-2.2) \\
J0051-7321	&	CC 	& 	5	&	22.3(+0.6,-0.5)	&	21.4(+1.3,-2.1)	&	79.6(+4.9,-8.1) \\
J0052-7236	&	CC 	& 	5	&	34.5(+10.9,-14.8)	&	46.7(+114.0,-30.8)		&10.5(+25.5	,-7.1) \\
J0056-7209	&	unknown 	& 	1.4	&20.6(+19.0,-10.6) &	22.4(+71.7,-16.3)	&	2.8(+2.6,-0.9) \\
J0057-7211	&	unknown 	& 	1.4	&10.3(+2.6,-2.9)	&	12.3(+11.5,-4.5)	&	5.5(+0.7,-0.8) \\
J0058-7217	&	CC 	& 	5	&	29.6(+4.4,-7.1)	&	36.2(+27.0,-18.4)	&	13.9(+10.4,-7.2) \\
J0059-7210	&	CC 	& 	5	&	12.1(+0.3,-0.3)	&	25.9(+1.3,-1.2)		&33.1(+1.5,-1.4) \\
J0100-7133	&	CC 	& 	5	&	20.9(+5.1,-3.0)	&	13.1(+7.0,-4.6)		&6.4(+3.8,-1.1) \\
J0103-7209	&	CC 	& 	5	&	15.6(+3.1,-3.6)	&	43.3(+30.4,-13.0)	&	3.4(+0.4,-0.4) \\
J0103-7247	&	CC 	& 	5	&	5.1(+0.4,-0.3)	&	8.7(+1.4,-1.4)		&10.6(+2.6,-2.4) \\
J0104-7201$^{a}$	&	CC 	& 	5	&4.2(+0.1,-0.3)	&7.2(+1.2,-0.8)&338(+43,-40) \\
J0105-7210	&	Ia? 	& 	1.4	&	12.4(+1.4,-0.2)	&	13.1(+2.0,-2.5)		&12.2(+2.1,-1.1) \\
J0105-7223	&	CC 	& 	5	&	30.8(+0.8,-0.8)	&	47.0(+2.6,-2.8)	&	83.0(+4.6,-4.8) \\
J0106-7205	&	Ia? 	&  	1.4	&	7.7(+0.5,-0.4)	&	9.8(+1.1,-1.2)	&	37.5(+2.8,-3.0) \\
J0127-7333	&	unknown 	& 	1.4	&26.7(+0.7,-1.3)	&	9.1(+2.9,-2.6)	&	25.0(+7.8,-7.1) \\
\enddata
\tablenotetext{a}{This is SNR E0102-7219. For ejecta masses of 20 and 40 $M_{\odot}$,
ages (in kyr) are 4.9(+0.1,-0.4) and 5.6(+0.1,-0.4), energies (in $10^{50}$ erg) are 6.7(+1.2,-0.9)  
and  6.5(+1.1,-0.4), and densities (in $10^{-2}$ cm$^{-3}$) are  318(+44,-40)
and 294(+40,-37).}
\end{deluxetable}

\clearpage

\begin{longrotatetable}
\begin{deluxetable*}{crrrrrrrrrrr} 
\tabletypesize{\scriptsize}

\tablecaption{SNRs with Two Components: Predicted Reverse Shock Properties} 
\tablewidth{0pt}
\tablehead{\\
\colhead{SNR/Type$^{a}$} & s,n & \colhead{Age(+.-)} &  \colhead{Energy(+,-)}    
& \colhead{$n_{-2}$(+,-) or $\rho_{s,14}$(+,-)}& \colhead{$kT_{r}$(+,-)} & \colhead{$EM_{r}$(+,-)} \\ 
\colhead{} & \colhead{} & \colhead{(kyr)} & \colhead{($10^{50}$erg)}  & {($10^{-2}$ cm$^{-3}$ or} & \colhead{(keV)} & \colhead {($10^{56}$cm$^{-3}$)} \\
\colhead{} & \colhead{} & \colhead{} & \colhead{} & \colhead{$10^{13}M_{\odot}$s/(km~yr))} &\colhead{} &\colhead{} 
}
\startdata
J0049-7314/Ia? Observed	&		&   n/a & n/a & n/a  & 0.91(0.03,-0.03)   & 728(+125,-99) \\
J0049-7314 Models &	0,7 	&   18.1(+3.5,-2.6) & 11.5(+4.7,-3.4) & $n_{-2}$=7.3(+3.3,-2.2)  & 3.6(1.1,-1.0)   & 0.11(+0.02,-0.02) \\
   &	0,10 	&   17.9(+3.4,-2.6) & 11.7(+4.9,-3.4)& $n_{-2}$=7.3(+3.3,-2.2)  & 2.2(+0.6,-0.6)   & 0.19(+0.03,-0.03) \\
         &	2,6	&   6.6(+2.5,-1.5) & 38(+26,-18)& $\rho_{s,14}$=2.05(+0.95,-0.63)  & 0.18(+0.05,-0.05)   & 156(+180,-81) \\
      &	2,7	&   1.63(+0.38,-0.24) & 218(+82,-74)& $\rho_{s,14}$=0.86(+0.40,-0.26)  & 0.55(+0.16,-0.15)   & 102(+117,-53) \\
            &	2,8	&   2.65(+0.77,-0.46) & 66(+30,-26)& $\rho_{s,14}$=1.14(+0.53,-0.35)  & 0.20(+0.06,-0.05)   &430(+490,-220) \\
            &	2,10	&   2.70(+0.78,-0.46) & 38(+17,-15)& $\rho_{s,14}$=1.08(+0.50,-0.33)  & 0.11(+0.03,-0.03)   & 1300(+1500,-700) \\
J0051-7321/CC Observed		&		&   n/a & n/a & n/a  & 0.73(0.16,-0.10)   & 1240(+526,-215) \\
J0051-7321  Models &	0,7 	& 22.3(+0.6,-0.5)&21.4(+1.3,-2.1)&	$n_{-2}$=79.6(+4.9,-8.1)& 1.74(+0.08,-0.10) &3.60(+0.09,-0.16)  \\
   &	0,10 	&   22.1(+0.6,-0.5) & 21.9(+1.4,-2.2)& $n_{-2}$=79.6(+5.0,-8.1)  & 1.08(+0.05,-0.07)& 6.16(+0.16,-0.27) \\
                  &	2,6	&  13.7(+0.4,-0.3) & 22.9(+1.2,-1.5)& $\rho_{s,14}$=13.2(+0.8,-1.4)  & 0.072(+0.003,-0.003)   & 8500(+1100,-1700) \\
         &	2,7	&   3.89(+0.20,-0.19) & 93(+9,-9)& $\rho_{s,14}$=5.5(+0.4,-0.6)  & 0.22(+0.01,-0.01)   & 5500(+700,-1200) \\
        &	2,8	&   7.61(+0.32,-0.29) & 18.7(+1.5,-1.5)& $\rho_{s,14}$=7.32(+0.46,-0.76)  & 0.080(+0.004,-0.004)   & 23400(+3100,-4600) \\
        &	2,10	&   7.74(+0.33,-0.30) & 10.5(+0.9,-0.8)& $\rho_{s,14}$=6.94(+0.44,-0.72)  & 0.043(+0.002,-0.002)   & 71000(+9000,-14000) \\
J0104-7201/CC Observed	 &		&   n/a & n/a & n/a  & 0.62(0.10,-0.10)   & 8100(+800,-800) \\
J0104-7201 Models &	0,7	&4.19(+0.08,-0.32)	&7.2(+1.2,-0.8)&$n_{-2}$=338(+43,-40)  & 1.03(+0.17,-0.11)& 40.4(+5.6,-4.5)\\
   &	0,10 	& 4.15(+0.08,-0.31)	&7.3(+1.3,-0.8)&$n_{-2}$=339(+44,-41)  & 0.58(+0.12,-0.11)& 61.2(+9.7,-8.9)\\
            &	2,6	& 2.22(+0.06,-0.25) & 18.8(+5.2,-0.9)& $\rho_{s,14}$=5.12(+0.64,-0.50)  & 0.200(+0.035,-0.007)   & 4270(+1240,-860) \\
            &	2,7 & 0.461(+0.014,-0.058) & 206(+64,-12)& $\rho_{s,14}$=2.15(+0.27,-0.25)  & 0.604(+0.105,-0.021)   & 2800(+740,-620) \\
            &	2,8	& 0.909(+0.029,-0.127) & 50.3(+19.5,-3.5)& $\rho_{s,14}$=2.85(+0.35,-0.34)  & 0.223(+0.039,-0.008)   & 11800(+3200,-2600) \\
 &	2,10	& 0.921(+0.034,-0.138) & 36.3(+14.0,-2.5)& $\rho_{s,14}$=2.71(+0.34,-0.32)  & 0.121(+0.021,-0.004)   & 36200(+9700,-8000) \\
J0105-7210/Ia? Observed	 &		&   n/a & n/a & n/a  & 0.78(0.07,-0.30)   & 188(+63,-58) \\
J0105-7210  Models &	0,7	& 12.4(+1.4,-0.2)&13.1(+2.0,-2.5)&$n_{-2}$=12.2(+2.1,-1.1)& 4.30(0.41,-0.70)  & 0.20(+0.01,-0.01)  \\
   &	0,10 	&   12.3(+1.4,-0.2) & 13.3(+2.0,-2.5)& $n_{-2}$=12.2(+2.1,-1.1)  & 2.64(+0.26,-0.43)& 0.35(+0.02,-0.01) \\
            &	2,6	&  4.28(+0.86,-0.13) & 51.0(+3.2,-15.6)& $\rho_{s,14}$=2.16(+0.38,-0.20)  & 0.241(+0.007,-0.040)   & 220(+85,-39) \\
            &	2,7	&   1.11(+0.13,-0.02) & 270(+17,-54)& $\rho_{s,14}$=0.91(+0.16,-0.08)  & 0.72(+0.02,-0.12)   & 143(+55,-25) \\
                        &	2,8	&   1.78(+0.26,-0.04) & 85.2(+3.8,-20.2)& $\rho_{s,14}$=1.20(+0.20,-0.11)  & 0.264(+0.008,-0.043)   & 608(+236,-107) \\
 &	2,10	&   1.81(+0.26,-0.04) & 50.2(+2.2,-11.8)& $\rho_{s,14}$=1.14(+0.20,-0.11)  & 0.144(+0.004,-0.023)   & 1850(+710,-330) \\
J0105-7223/CC Observed	&		&   n/a & n/a & n/a  & 0.38(0.01,-0.01)   & 8530(+2630,-2070) \\
J0105-7223 Models &	0,7 	& 30.8(+0.8,-0.8)&47.0(+2.6,-2.8)	&$n_{-2}$=83.0(+4.6,-4.8)& 2.66(+0.14,-0.14) & 2.23(+0.05,-0.05)  \\
   &	0,10 	&   30.6(+0.8,-0.8) & 48.1(+2.7,-2.8)& $n_{-2}$=83.1(+4.6,-4.9)  & 1.65(+0.09,-0.09)& 3.81(+0.08,-0.09) \\
               &	2,6	&  19.2(+0.5,-0.5) & 37.4(+2.0,-2.0)& $\rho_{s,14}$=24.5(+1.4,-1.5)  & 0.065(+0.003,-0.003)   & 22200(+2600,-2600) \\
               &	2,7	&   6.24(+0.34,-0.31) & 100(+11,-10)& $\rho_{s,14}$=10.3(+0.6,-0.6)  & 0.20(+0.01,-0.01)   & 14400(+1700,-1700) \\
                              &	2,8	&   11.50(+0.47,-0.44) & 20.6(+1.7,-1.6)& $\rho_{s,14}$=13.6(+0.8,-0.8)  & 0.071(+0.004,-0.004)   & 61300(+7100,-7100) \\
                &	2,10	&  11.7(+0.5,-0.5) & 10.4(+0.9,-0.8)& $\rho_{s,14}$=12.9(+0.7,-0.8)  & 0.039(+0.002,-0.002)   & 1.8$\times10^5$(+2$\times10^4$,-2$\times10^4$) \\
J0106-7205/Ia Observed	&		&   n/a & n/a & n/a  & 1.00(0.03,-0.03)   & 788(+103,-95) \\
J0106-7205/Ia Models &	0,7	&  	7.7(+	0.5,-0.4)	&9.8(+1.1,-1.2)	&	$n_{-2}$=37.5(+2.8,-3.0)& 3.83(+0.38,-0.42)  & 0.74(0.03,-0.03)  \\
   &	0,10 	&   7.6(+0.5,-0.4) & 9.9(+1.1,-1.2)& $n_{-2}$=37.4(+2.8,-2.9)  & 2.35(+0.23,-0.26)& 1.28(+0.04,-0.04) \\
                  &	2,6	&  2.97(+0.37,-0.27) & 33.2(+7.0,-7.0)& $\rho_{s,14}$=2.55(+0.20,-0.20)  & 0.254(+0.025,-0.029)   & 492(+79,-75) \\
                &	2,7	&   0.70(+0.06,-0.04) & 227(+29,-33)& $\rho_{s,14}$=1.07(+0.08,-0.08)  & 0.76(+0.07,-0.08)   & 320(+51,-49) \\
                                &	2,8	&   1.16(+0.12,-0.08) & 69(+11,-12)& $\rho_{s,14}$=1.42(+0.11,-0.11)  & 0.280(+0.027,-0.031)   & 1360(+220,-210) \\
                &	2,10	&  1.18(+0.12,-0.08) & 42.2(+6.7,-7.4)& $\rho_{s,14}$=1.35(+0.10,-0.11)  & 0.153(+0.015,-0.017)   & 4140(+660,-630) \\
\enddata
\label{tab:customS2}
\tablenotetext{a}{For CC-type, we use $M_{ej}=5M_{\odot}$; for Ia-type, we use $M_{ej}=1.4M_{\odot}$; for both we use CC type abundances.}
\end{deluxetable*}
\end{longrotatetable}

\clearpage

\begin{deluxetable}{cccccccc}
\tabletypesize{\scriptsize}
\tablecaption{Comparison of SMC, LMC and Galactic SNR Rates, Energies and Densities} 
\tablewidth{0pt}
\tablehead{\\
\colhead{System} & \colhead{Mean SNR Rate} &  \colhead{2$\sigma$ Rate Range} &  \colhead{Mean Energy}   &  \colhead{1$\sigma$ Energy Dispersion} & \colhead{Density}  & \colhead{1$\sigma$ Density Dispersion}\\
\colhead{}            & \colhead{(1/yr)}          & \colhead{(1/yr)}            & \colhead{($10^{51}$ erg)}  & \colhead{(ratio)}                                   & \colhead{( cm$^{-3}$)} & \colhead{(ratio)} 
}
\startdata
SMC 20 SNRs$^{a}$	&	1230 	&  1100-1350 & 1.6 &	1.87	&	0.14  & 3.06\\
LMC 50 SNRs$^{b}$	&	503 	&  490-525     & 0.48 &	2.94	&	0.079  &  3.24\\
Galaxy 15 SNRs (l=18-54$^\circ$)$^{c}$	&   870 & 740-1000       & 0.54 &	3.47	&	0.26  &  6.31\\
Galaxy 43 SNRs (l=38-360$^\circ$)$^{d}$ &	324	&  302-347     & 0.27 &	3.47	&	0.069  &  5.13\\
\enddata
\label{tab:stats}
\tablenotetext{a}{This work.}
\tablenotetext{b}{SNR ages, energies and densities from \cite{2017Leahy}, error analysis this work.}
\tablenotetext{c}{SNR ages, energies and densities from \cite{2018ApJ...866....9L},  error analysis this work.}
\tablenotetext{d}{SNR ages, energies and densities from \cite{2020ApJS..248...16L},  error analysis this work.}
\end{deluxetable}

\end{document}